\documentstyle[prl,twocolumn,epsfig,aps]{revtex}
 
\newcommand{\be}{\begin{equation}}
\newcommand{\ee}{\end{equation}}
\newcommand{\bea}{\begin{eqnarray}}
\newcommand{\eea}{\end{eqnarray}}
\newcommand{\no}{\noindent}

\begin{document}
\draft

\title{A study of the 't~Hooft loop in $SU(2)$ Yang--Mills theory }
\author{Philippe de Forcrand$^{1,2}$, Massimo D'Elia$^3$ and Michele Pepe$^1$}
\address{$^1$Institut f\"ur Theoretische Physik, ETH H\"onggerberg,
  CH-8092 Z\"urich, Switzerland}
\address{$^2$CERN, Theory Division, CH-1211 Gen\`eve 23, Switzerland}
\address{$^3$Dipartimento di Fisica dell'Universit\`a and I.N.F.N., I-56127, Pisa, Italy}

\date{\today}
\maketitle
 
\begin{abstract}
We study the behaviour of the spatial and temporal 't~Hooft loop at zero and 
finite temperature in the $4D$ $SU(2)$ Yang-Mills theory, using a new numerical 
method. In the deconfined phase $T>T_c$, the spatial 't~Hooft loop exhibits
a dual string tension, which vanishes at $T_c$ with
$3D$ Ising-like critical exponent.
\end{abstract}
 

The $4D$ $SU(2)$ Yang-Mills theory undergoes a transition between a
cold confined phase and a hot deconfined phase at a critical temperature $T_c$. 
An order parameter widely used to characterize this transition is the Polyakov
loop. It develops a non-vanishing expectation value in the deconfined
phase. However, the corresponding operator creates a single fundamental
static color source, which does not belong to the physical Hilbert
space of the theory; 
it cannot be defined at zero temperature; and it is afflicted by ultra-violet
divergences in the continuum limit \cite{Smilga,Kaltes2}.
A long time ago it was proposed \cite{tHooft1} to consider instead the
't~Hooft loop operator as an order parameter to characterize 
this transition. It is the purpose of this paper to study this order parameter.
The 't~Hooft loop, $\tilde{W}(C)$, is an operator associated with a
given closed contour $C$, and is defined in the continuum $SU(N)$ theory by the 
following equal--time commutation relations\cite{tHooft1}
\bea
\left[ W(C), W(C') \right]~=~\left[ \tilde{W}(C), \tilde{W}(C') \right]~=~0~ \\
\tilde{W}^\dagger(C) W(C') \tilde{W}(C)~=~e^{i \frac{2\pi}{N} n_{CC'}} W(C')
\eea
where $W(C')$ is the Wilson loop associated with the closed contour $C'$ and
$n_{CC'}$ is the linking number of $C$ and $C'$. 
Just like the Wilson loop creates 
an elementary electric flux along $C'$,
the 't~Hooft loop
creates an elementary
magnetic flux along the closed path $C$ 
affecting any Wilson loop ``pierced'' by $C$. 
In that sense, the two types of loop are dual to each other. 
At zero temperature, it has been shown \cite{tHooft1,Tomb,Sam} 
that also the 't~Hooft loop behaviour is dual to that of the Wilson loop:
in the absence of massless excitations, an area law behaviour for one implies 
a perimeter law for the other, and vice versa. 
Hence, at $T=0$ the 't~Hooft loop obeys a perimeter law.

Several analytical \cite{Kaltes2,Kaltes1} and numerical \cite{KandT,Rebbi,Delde}
studies have been carried out in order  
to investigate this issue of duality at finite temperature. At
$T>0$, the Lorentz symmetry is broken, so spatial and 
temporal loops can have different behaviours. Because the 
spatial string tension persists also above $T_c$ for the Wilson loop, 
temporal 't~Hooft loops are expected to show a perimeter law in both phases; 
spatial 't~Hooft loops are expected to obey a perimeter law in the confined 
phase and an area law -- defining a dual string tension (strictly
speaking it is an action density) -- in the 
deconfined phase. 

On the lattice the 't~Hooft loop is defined as follows.  
Let us consider the $SU(2)$ lattice gauge theory with the usual
Wilson plaquette action,
\be
S(\beta) = \beta \sum_P (1 - \frac{1}{2} \mbox{Tr} (U_P))\; ,
\label{wilsact}
\ee
where the sum extends over all the plaquettes $P$, and $U_P$ is the
path--ordered product of the links around $P$. Starting from $S(\beta)$,
 one defines the partition function 
\be
Z(\beta) = \int [dU] \exp(- S(\beta))\; .
\label{wilspfc}
\ee
Let us now switch on ``by hand'' an elementary magnetic flux along a
closed contour $C$ defined on the dual lattice. 
To create this
magnetic flux, we have to multiply $U_P$ by 
a non-trivial element of the center group for the plaquettes $P$ dual
to a given surface ${\cal{S}}$ supported by $C$\cite{Guth,Suss}.
For the $SU(2)$ gauge group, this means flipping the
sign of the coupling, since the only non-trivial element of the center $Z_2$ is
$-1$. Call $\cal P(\cal S)$ the 
set of plaquettes whose coupling is flipped $\beta \to - \beta$.
Then the action of the system where an elementary flux along a closed
contour $C$ has been switched on is given, up to an additive constant, by
\be
S_{\cal S}(\beta) = - \frac{1}{2} \beta \left( 
\sum_{P \notin {\cal P(\cal S)}}\mbox{Tr} (U_P) - 
\sum_{P \in {\cal P(\cal S)}}\mbox{Tr} (U_P) 
                                                        \right) \; ,
\label{modiact}
\ee
and the partition function is
\be
Z_C(\beta) = \int [dU] \exp(- S_{\cal S}(\beta))\; .
\label{modipfc}
\ee
$Z_C(\beta)$ does not depend on the particular chosen
surface $\cal S$, since different choices are related by a 
change of integration variables. The simplest choice for 
$\cal S$ is the minimal surface spanning $C$. Thus, if $C$ is
an $R_x \times R_y$ rectangle in the $(x,y)$ plane (spatial 't~Hooft loop),
one flips the coupling of the $(z,t)$ plaquettes dual to the plaquettes 
belonging to the rectangular area.

The 't~Hooft loop expectation value is related to the free energy cost 
needed to create the magnetic flux along the contour $C$ and is given by  
\be
\langle \tilde{W}(C) \rangle = 
Z_C(\beta) / Z(\beta) \;.
\label{tllatdef}
\ee
This expression can be rewritten in the form
\be
\langle \tilde{W}(C) \rangle = \langle \exp 
\big( - \beta \sum_{P \in {\cal P(\cal S)}}\mbox{Tr} (U_P) \big)\rangle \; ,
\label{tlprac}
\ee
with the average taken with the standard Wilson action.
In this form, the difficulty of measuring the 't~Hooft loop
becomes clear: the observable
is exponentially suppressed on typical configurations
of the statistical ensemble, and gets a significant contribution only from  
configurations having an extremely small statistical weight.
Therefore the numerical evaluation of $\langle \tilde{W}(C) \rangle$
represents a difficult sampling problem, increasingly so with the 
loop size.

Recently, the 't~Hooft loop, or special cases of it, have been studied
numerically on the lattice. In \cite{KandT,Rebbi} the sampling problem 
was overcome by using a multihistogram method, where one performs several
different simulations in which the coupling associated with the plaquettes
in $\cal P(S)$ is gradually changed from $\beta$ to $-\beta$.
In \cite{Delde}, instead, the derivative 
${\rm d}/{\rm d}\beta \ln \langle \tilde{W}(C) \rangle$ has been determined,
which is a much simpler numerical task.

In this letter we report on a similar numerical study where, by measuring
spatial and temporal 't~Hooft loops at zero and
finite temperature, we confirm the role of the 't~Hooft loop
as a dual order parameter for confinement. We adopt
a new numerical method, rewriting the
ratio  $Z_C(\beta) / Z(\beta)$ as a product of intermediate
ratios, each easily measurable.
We establish the perimeter law behaviour of the
't~Hooft loop at zero temperature. We measure the free energy of two
center monopoles as a function of their separation: 
as expected from duality,
temporal 't~Hooft loops show screening behaviour at all temperatures,
while spatial 't~Hooft loops exhibit a dual string tension 
above $T_c$. 
Moreover, this dual string tension
vanishes at $T_c$ with a critical exponent $\nu$ very close 
to that of the $3D$ Ising model, consistent with universality.

{\it The numerical technique} -- We now describe our numerical method to measure
the expectation value of the 't~Hooft operator. 
Because the relevant contributions to $Z_C(\beta)$ and to $Z(\beta)$ come from
regions of the phase space with very poor overlap, a direct evaluation 
of  (\ref{tllatdef}), (\ref{tlprac}) by a single Monte Carlo simulation is not 
a reliable way to compute $\langle \tilde{W}(C) \rangle$. 
It is necessary
to consider a sequence of intermediate partition functions which interpolate 
between $Z(\beta)$ and $Z_C(\beta)$.
The approach used in \cite{KandT,Rebbi} consists of interpolating in the {\em coupling} of the flipped plaquettes in $\cal P(S)$, from $\beta$ to $-\beta$.
We interpolate instead in the {\em number} of flipped plaquettes,
and rewrite $Z_C(\beta) / Z(\beta)$ as the
product of ratios of partition functions where the number of plaquettes with
flipped coupling is progressively reduced. $N$ being the number of
plaquettes in $\cal P(S)$, we use the identity 

\be
\frac{Z_C(\beta)}{ Z(\beta)} =
\frac{Z_N(\beta)}{ Z_{N-1}(\beta)} \cdot
\frac{Z_{N-1}(\beta)}{ Z_{N-2}(\beta)} \cdot\ldots
\cdot \frac{Z_{1}(\beta)}{ Z_0(\beta)}
\label{factor}
\ee

\noindent
where $Z_k(\beta)$, $k=0,\dots , N$ ($Z_N\equiv Z_C$ and $Z_0 \equiv
Z$) is the partition function of the system
where only the first $k$ plaquettes in $\cal P(S)$ have flipped
coupling. Every ratio $Z_k/Z_{k-1}$ can be now
computed efficiently by a single Monte Carlo simulation, due to the 
good overlap of the relevant phase space in the two partition
functions. Furthermore, in the practical implementation, it is useful to reexpress 
$Z_k (\beta)/Z_{k-1} (\beta )$ as a ratio of expectation values

\be
\frac{Z_k (\beta)}{Z_{k-1} (\beta )} = 
\frac{\langle \exp(-\frac{1}{2}\beta\mbox{Tr}(U_{P_k})) \rangle_k}
{\langle \exp(+\frac{1}{2}\beta\mbox{Tr}(U_{P_k}) )\rangle_k}
\label{ratio}
\ee

\noindent
The averages $\langle \cdot \rangle_k$ are computed with respect to an
action where the $k$th plaquette $U_{P_k}$ of 
$\cal P(S)$ has zero coupling, the first $(k-1)$ plaquettes have
coupling $-\beta$ and the remaining ones coupling $+\beta$.
The benefits over simply computing
$\langle \exp(-\beta\mbox{Tr}(U_{P_k}) )\rangle$
using the distribution corresponding to $Z_{k-1} (\beta )$ are two-fold.
$(i)$ Using the intermediate weight, $\langle \cdot \rangle_k$, 
allows a much better sampling for both 
$\langle \exp(-\frac{1}{2}\beta\mbox{Tr}(U_{P_k})) \rangle_k$
and
$\langle \exp(+\frac{1}{2}\beta\mbox{Tr}(U_{P_k})) \rangle_k$,
with reduced errors.
$(ii)$ Since the quantity $Z_k (\beta) / Z_{k-1} (\beta )$ 
refers to a single plaquette of the lattice, it
is very useful to perform a partial integration
after each updating sweep of the whole lattice,
and measure
$\exp(+\frac{\beta}{2}\mbox{Tr}(U_{P_k}))$ and
$\exp(-\frac{\beta}{2}\mbox{Tr}(U_{P_k}))$
several times
while updating only the four links belonging to plaquette
$P_k$. Since this plaquette has zero coupling, 
these four links are decoupled from each other. Therefore
each link can be updated independently $N_{\rm hit}$ times
and the different copies of each link can
be combined to obtain $N_{\rm hit}^4$ measurements of 
$\exp(\pm\frac{\beta}{2}\mbox{Tr}(U_{P_k}))$.
Although these measurements are correlated, the variance reduction
is important.

The statistical error on ratio
(\ref{ratio})
must be evaluated with care, since the two averages are computed from the 
same sample of configurations. We use a jackknife analysis.
Then, after each $Z_k (\beta) / Z_{k-1} (\beta )$ has been computed,
the 't~Hooft loop expectation value is evaluated according to
Eq.(\ref{factor}). The final statistical error is obtained
by standard error propagation since each  
$Z_k (\beta) / Z_{k-1} (\beta )$ comes from an independent Monte Carlo 
simulation.

An advantage of our method over the multihistogram technique \cite{KandT,Rebbi}
is that the products in (\ref{factor}) give us information on
smaller 't~Hooft loops for free; moreover, the error analysis is simpler
and less delicate than in a multihistogram analysis.

{\it Results} -- We focus on the free energy $F(R)$ of a pair of static 
center monopoles as a function of their separation $R$. It can be obtained as 
${\rm lim}_{R_t \rightarrow \infty} -{\rm Ln}[\langle \tilde{W}(R,R_t) \rangle] / R_t$
by taking elongated $R \times R_t$ rectangular loops, in the same way as one
extracts the static potential between two chromo-electric charges. We take $R_t$
as large as possible, i.e. equal to the lattice size $L$.
This is analogous to measuring the correlation of two Polyakov loops, and is
the correct approach at finite temperature. Therefore we must flip the
coupling of $R \times L$ plaquettes. We do this by scanning first the $R_t-$,
then the $R-$direction. 
With this ordering, the intermediate partition functions
$Z_L, Z_{2L},..,Z_{R \times L}$ in (\ref{factor})
provide us with the free energy at separations $1, 2,.., R$ respectively.
The final ratio $Z_{L \times L}/Z_0$ gives the free energy of a center vortex
as computed in \cite{KandT}.
Our lattice sizes range from $10^3 \times 2$ to $20^3 \times 10$, at couplings
$\beta=\frac{4}{g^2}$ from $2.3$ to $2.8$, for which the lattice
spacing varies by a factor $\sim 5$. We perform 5-10000
(multi-hit) measurements of each ratio (\ref{ratio}).

\begin{figure}[h]
\begin{center}
\epsfig{figure=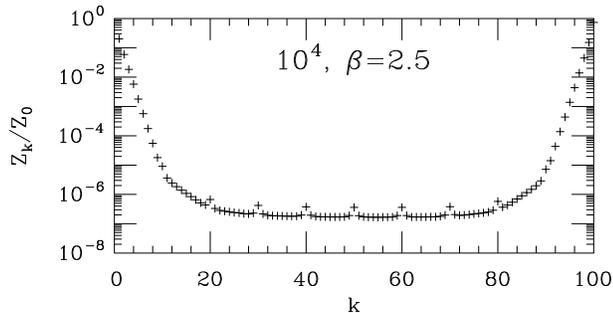,height=3.5cm,width=7cm}
\end{center}
\caption{Partition function $Z_k/Z_0$ versus $k$. Incrementing $k$ always
increase by $1$ the area of the 't~Hooft loop, but only changes its
perimeter by $\mp 2$ if $k = 0$ or $1$ mod(10) respectively. $Z_k$ is clearly 
sensitive to changes of perimeter but not of area.}
\label{perimeter}
\vspace{-0.1cm}
\end{figure}

At zero temperature the 't~Hooft loop is expected to obey a perimeter law:
$Z_k / Z_0 \propto e^{-c \tilde{P}_k}$, where $\tilde{P}_k$ is the length of the contour
$C_k$. For the sequence of flipped plaquettes defined above, 
$\tilde{P}_k = 2 L$ if ${\rm mod}(k,L) = 0$, $(2 L + 2)$ otherwise, unless 
$k < L$ or $k > L(L - 1)$. $Z_k$ should therefore center around two values
only. This is exactly what appears in Fig.1, on a $10^4$ lattice.
The free energy is clearly insensitive to changes in the 't~Hooft loop area.
This is also confirmed by a direct measurement of Creutz ratios
$\chi(R,R)$, which estimate the force at distance $R$ between the two
magnetic charges and quickly drop to zero as $R$ increases.
As with the Wilson loop,
the coefficient $c$ of the perimeter term is affected by UV divergences.
We verified that it increases as the
lattice spacing is reduced.

The free energy $F(r)$ can be fitted by a Yukawa form $\frac{e^{-m r}}{r}$,
up to an irrelevant additive constant. However the screening mass $m$ is
rather large, so that the signal quickly dies out. Moreover, consistency of
data at different values of the lattice spacing is only fair. This is
presumably caused by short-distance lattice distortions of the Yukawa potential,
which could be explicitly taken into account. In any case, Fig.2
shows the free energy as a function of the magnetic charge separation.
The curve corresponds to a screening mass of $2$ GeV. It makes sense
to match this mass with the lightest gluonic excitation, the
scalar glueball ($\sim 1.65$ GeV), as attempted in \cite{Rebbi,Delde}; but
this issue awaits a more extensive numerical study.

\begin{figure}[th]
\begin{center}
\epsfig{figure=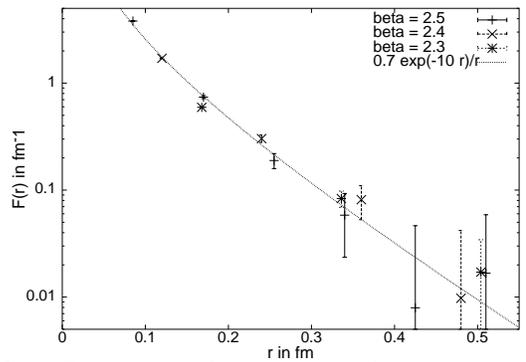,height=4.5cm,width=7cm}
\end{center}
\caption{Free energy of a static pair of center monopoles as a function
of their separation, at temperature $T < T_c$. The scaling behaviour
of the data is mediocre. The curve shows a Yukawa
potential with screening mass $2$ GeV.}
\label{screeningT=0}
\vspace{-0.1cm}
\end{figure}

At finite temperature we must distinguish between spatial and temporal
't~Hooft loops. 
In the case of electric charges, the spatial string tension
persists above $T_c$, while the correlation of time-like Polyakov loops
develops a disconnected part showing saturation of the static potential.
Here the temporal 't~Hooft loop should show screening, while the
spatial 't~Hooft loop should obey an area law. It is clear
that the latter will cost more in free energy, since the center vortex
created by each flipped $(z,t)$ plaquette can only spread over a limited
time extent $T^{-1}$.

\begin{figure}[h]
\begin{center}
\epsfig{figure=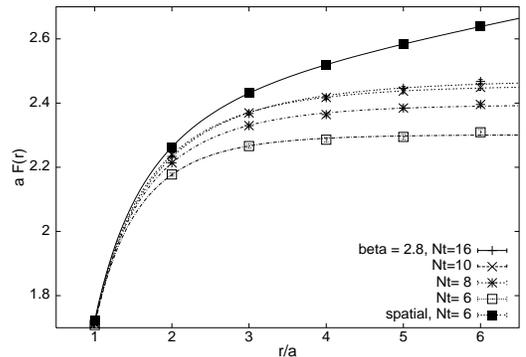,height=4.5cm,width=7cm}
\end{center}
\caption{Free energy of a static pair of center monopoles as a function
of their separation at temperature $T > T_c$. 
The spatial 't~Hooft loop shows a dual string tension;
the temporal 't~Hooft loop 
shows a screening mass which increases with $T$.}
\vspace{-0.1cm}
\end{figure}

Indeed, this is precisely what we observe. Fig.3 shows the same Yukawa form
for a temporal 't~Hooft loop at all temperatures. The only effect of 
temperature is to increase the screening mass. 
The fitting coefficient of the Yukawa potential is smaller than the 
short-distance perturbative prediction $1/4\pi (2 \pi/g)^2$ \cite{Poly}, but
grows towards this value at higher $\beta$.
An attempt to fit the data with the ansatz $F_0 + c \frac{e^{-m r}}{r} + \sigma r$,
including a linear term, gives a dual string tension $\sigma$ consistent with zero. 
In contrast, this linear term is required to obtain an acceptable fit
above $T_c$ for the spatial 't~Hooft loop: a dual string tension appears.
As a practical consequence, the screening mass becomes yet harder to determine.
It seems little affected by temperature, as Fig.4 (top) shows, unlike
the temporal screening mass (bottom) which rises more or less linearly
with $T$, much like the glueball excitation which it presumably represents
\cite{Datta}.
Precise quantitative statements about these dependences on $T$ require a more accurate
numerical investigation.

\begin{figure}[th]
\begin{center}
\epsfig{figure=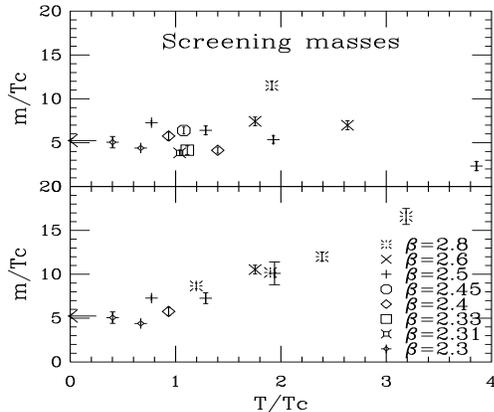,height=5.0cm,width=6.0cm}
\end{center}
\caption{Screening mass as a function of temperature, both in units of $T_c$,
as extracted from spatial (top) or temporal (bottom) 't~Hooft loops.
Below $T_c$ both coincide.
The arrow gives the mass of the scalar glueball at $T=0$.}
\label{screening_mass}
\vspace{-0.1cm}
\end{figure}

The dual string tension $\sigma$ depends on temperature and must vanish
at $T_c$. Fig.5 shows that it does so as 
$\sigma \propto \linebreak \left(\frac{T-T_c}{T_c}\right)^{2\nu}$.
The critical exponent $\nu$, associated with the correlation length
$\xi = \sigma^{-1/2}$, comes out very close to that of the $3D$
Ising model: $0.66(3)$ vs. $\approx 0.63$.
Indeed this should be expected since both models are in the same 
universality class, although the Ising exponent is typically extracted
from the divergence of $\xi$ in the {\em symmetric} (confined) phase.
This dual string tension can then be taken as order parameter for the
restoration of the (magnetic) $Z_N$ symmetry, corresponding to 
{\em de}confinement \cite{Kaltes2,Kovner}.

Fig.5 also shows high $T$, perturbative results.
Notice that a spatial $(x,y)$ 't~Hooft loop of maximal size $L \times L$
introduces a flipped plaquette in every $(z,t)$ plane, which is equivalent
to enforcing twisted boundary conditions in the $(z,t)$ directions,
thus creating a $Z_N$ interface with tension $(\sigma T)$. This interface tension
has been calculated to two loops \cite{Kaltes1}. Taking the running
coupling $g(T)$ from \cite{Kajantie}, one obtains the two curves in Fig.5
for the leading and next order. The numerical data lie in between.

In summary, by using a dual observable, we have measured a dual string 
(or interface)
tension in the deconfined phase. The corresponding correlation length 
diverges at $T_c$ with the $3D$ Ising critical exponent $\nu$ as expected 
from universality. 
An extension to $SU(3)$ is straightforward. In that case, since the
transition is first-order, the dual string tension will persist all
the way to $T_c$. 
Its value at $T_c$ has been the object of many studies \cite{Iwasaki}. 
Finally, it would be very desirable to measure more precisely the screening
mass and to clarify its physical origin.

\begin{figure}[t]
\begin{center}
\epsfig{figure=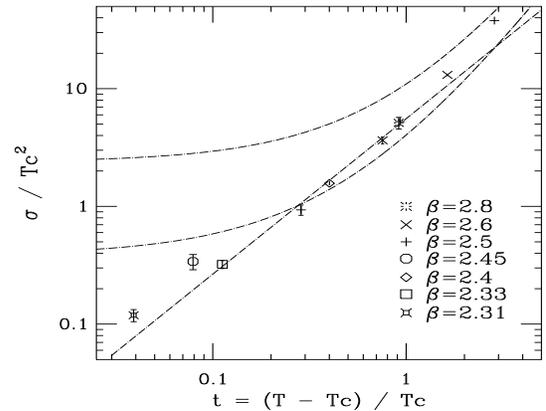,height=5.0cm,width=6.5cm}
\end{center}
\caption{Dual string tension, in units of $T_c^2$, as a function of 
the reduced temperature $t$. The straight line is a power law fit to $t < 1$.
The fitted exponent is $1.32(6)$, to be compared with $2 \nu \approx 1.26$
for the $3D$ Ising model. The curves show the perturbative result,
to leading (upper) and next (lower) order.}
\label{sigma}
\vspace{-0.5cm}
\end{figure}

\par\bigskip
    
\no {\bf Acknowledgments:}
We thank C. Alexandrou, P. Butera, L. Del Debbio, A. Di Giacomo, 
C. Korthals--Altes, M. Laine and  O. Philipsen for
many helpful discussions, and acknowledge
communication with T. Kov\'acs, T. Tomboulis and C. Rebbi.

\end{document}